\documentstyle[preprint,aps,epsf]{revtex}
\newif\iftightenlines\tightenlinesfalse
\tightenlines\tightenlinestrue

\def\del{\delta}
\def\eslt{\not\!\!{E_T}}
\def\to{\rightarrow}

\def\te{\tilde e}

\def\tu{\tilde u}

\def\tst{\tilde t}
\def\ttau{\tilde \tau}
\def\tmu{\tilde \mu}
\def\tg{\tilde g}
\def\tnu{\tilde\nu}
\def\tell{\tilde\ell}

\def\tw{\widetilde W}
\def\tz{\widetilde Z}
\begin{document}
\draft
\preprint{\vbox{\baselineskip=14pt%
   \rightline{FSU-HEP-000927}
   \rightline{UH-511-974-00}
}}
\title{Can Precision Measurements of Slepton Masses Probe Right Handed
Neutrinos?}
\author{Howard Baer$^1$, Csaba Bal\'azs$^2$, J.~Kenichi Mizukoshi$^2$ and
Xerxes Tata$^2$}
\address{
$^1$Department of Physics,
Florida State University,
Tallahassee, FL 32306 USA
}
\address{
$^2$Department of Physics and Astronomy,
University of Hawaii,
Honolulu, HI 96822, USA
}
\date{\today}
\maketitle
\begin{abstract}
In a supersymmetric model, the presence of a right handed neutrino with
a large Yukawa coupling $f_{\nu}$ would affect slepton masses via its
contribution to the renormalization group evolution between the grand
unification and
weak scales. Assuming a hierarchichal pattern of neutrino masses, these
effects are large for only the third generation of sleptons. We
construct mass combinations to isolate the effect of $f_{\nu}$ from mass
corrections already expected from tau Yukawa couplings. 
We then
analyze the size of these effects, assuming that the
Super-Kamiokande data constrain 0.033~eV $\alt m_{\nu_{\tau}} \alt 0.1$~eV
and that neutrino masses arise via a see-saw mechanism. We also
explore whether these effects might be detectable in experiments at
future $e^+e^-$ linear colliders.  We find that $m_{\tnu_{\tau}}$ needs to be
measured with a precision of about 2-3\% to measure the effect of
$f_{\nu}$ if the neutrino and top Yukawa couplings unify at the grand
unification scale. In a simple case study, we find a precision of only
6-10\% might be attainable after several years of operation. If
the neutrino Yukawa coupling is larger, or in more complicated models of
neutrino masses, a determination of $\ttau_1$ and $\tnu_{\tau}$  masses
might provide a signal of a Yukawa interaction of neutrinos. 

\end{abstract}

\medskip


\section{Introduction and Motivation}

Supersymmetry\cite{reviews} is a leading candidate for physics beyond
the Standard Model (SM). The Minimal Supersymmetric Standard Model
(MSSM) posits a spin zero partner for each SM matter field, and spin
half partners for the gauge and Higgs fields. While all
the dimensionless couplings of the MSSM are completely determined in
terms of known SM couplings, the intractably large number of
dimensionful soft SUSY breaking parameters is usually reduced by
supplementing the MSSM by simple boundary conditions for their
renormalization group (RG) evolution\cite{rge}. These boundary
conditions are chosen so that flavour changing neutral currents (which
are generically large in a theory with many scalars) are suppressed to
acceptable levels.  The boundary conditions are determined by assumptions
about the new physics (concerning mediation of SUSY breaking) at a scale
between $\sim 10^3$~TeV and $M_{Planck}$. Various models
(mSUGRA, low energy gauge-mediated SUSY breaking,
anomaly-mediated SUSY breaking, gaugino-mediated SUSY breaking), each
leading to a characteristic pattern of MSSM soft breaking parameters,
have been proposed and analyzed.

Within the MSSM framework, neutrinos are massless and have no Yukawa
couplings. Evidence from the Super-Kamiokande experiment\cite{superk}
strongly suggests that there are neutrino oscillations and, almost
certainly, a neutrino mass. The favoured interpretation of these data is
$\nu_{\mu}-\nu_{\tau}$ oscillations, with $\Delta m^2 \sim 3\times
10^{-3}$~eV$^2$ and near-maximal mixing.  The see-saw
mechanism\cite{seesaw} provides an elegant way of giving small masses to
neutrinos: for each massive neutrino, one needs to introduce a SM gauge
singlet chiral superfield ${\hat N}^c$, whose fermion component is the
left handed anti-neutrino. A mass term in the superpotential leads to a
Majorana mass $M_N$ for the singlet neutrino as well as a lepton-number
violating mass for its scalar partner. This mass may be very large
and is expected to be $\sim M_{GUT}$, or even $M_{Planck}$ if the right
handed neutrino (RHN) is a singlet of the unifying gauge group. The
neutrino also acquires a Dirac mass $m_D$, proportional to the neutrino
Yukawa coupling $f_{\nu}$, from electroweak symmetry breaking. The
physical neutrino mass, which is obtained by diagonalizing the neutrino
mass matrix, is then $\simeq m_D^2/M_N$, and can be very small since
$m_D$ is comparable to or smaller than the electroweak scale.

Without further assumptions, the neutrino Yukawa coupling is
arbitrary. However, in models where top and neutrino Yukawa couplings
unify at some high scale\footnote{Yukawa unification would occur in
$SO(10)$ models since all the matter fields including the RHN belong to
a single representation. In the minimal $SO(10)$ model where all the
fermions get their mass from one Higgs multiplet, all Yukawa couplings
are unified. In more complicated models where the masses of up and down
type fermions arise from different Higgs fields, we have unification of
top and neutrino Yukawa couplings, and separately unification of bottom and
tau Yukawa couplings.} we expect $f_{\nu} \simeq f_t$, so that the
physical neutrino mass is $\simeq m_t^2/M_N$. Renormalization
effects\cite{parida} modify this by a factor $O(1)$, but do not change
the order of magnitude of the predicted neutrino mass.
These effects are accounted for in our analysis.

The introduction of the RHN\cite{pheno} and its associated Yukawa
coupling would change the predictions for $\tell_L$ and $\tnu_L$ masses
via new contributions to their RG evolution, but would not (at one loop)
affect $\tell_R$ or other sparticle masses. Moreover, if we assume that
neutrino masses are hierarchical (as they must be in the GUT model with
the simplest see-saw mechanism), the $\Delta m^2$ measured by the
Super-Kamiokande collaboration is essentially $m_{\nu_{\tau}}^2$, with
the masses of other neutrinos being negligible in comparison. In the
rest of this analysis we will therefore assume that, like quark and
lepton Yukawa couplings, the neutrino Yukawa coupling of just the third
generation is relevant. Since this new coupling can affect just
$\ttau_L$ and $\tnu_{\tau}$ masses, precision measurements of these
masses in future experiments at $e^+e^-$ linear colliders (relative to
$\ttau_R$ and first generation slepton masses) should yield information
about $f_{\nu}$. An examination of how large these effects might be, and
an evaluation of the whether they might be detectable in the future,
forms the subject of this study.\footnote{While this paper was under
preparation, we became aware that the same idea had already been
suggested in Ref.\cite{moroi}.  This study did not include
intra-generational mixing which spoils the simple picture (see below),
and also neglected Yukawa couplings in the charged lepton as well as
down type quark sector (the latter are eliminated from our
considerations). Finally, our examination goes beyond Ref.\cite{moroi}
in that we also examine whether these effects might actually be
measurable in future experiments.}

The conceptually simple picture is spoiled by two important effects that
are already present in the MSSM. First, the usual tau Yukawa interactions
already split stau masses from those of other sleptons, and the effects
of these depend on the unknown parameter $\tan\beta$ whose value is
difficult to pin down except under special circumstances ({\it e.g.}
when the masses of several Higgs bosons can be
measured\cite{barger}). Second, the same Yukawa interactions also cause
$\ttau_L-\ttau_R$ mixing, so that $m_{\ttau_L}$ and $m_{\ttau_R}$ are not
the physical masses of the staus.

In the next section, we identify slepton mass combinations that are
sensitive to $\nu_{\tau}$ Yukawa couplings but not to the usual tau
Yukawa interactions, and compare expectations for these within a
reference model (mSUGRA) and the same model extended by a RHN.  In
Section~III, we perform a case study to estimate the precision with
which these variables might be measured at linear colliders in order to
assess whether such measurements might be feasible. We conclude in
Sec. IV with a summary of our results and some general remarks about how
our conclusions depend on the underlying model for neutrino masses.

\section{Isolating the effects of the Neutrino Yukawa Coupling}

Although we will limit ourselves to analyzing the effects of adding a
RHN to the mSUGRA framework, the ideas are applicable to any model with
a well-defined prediction for slepton masses such as gauge-mediated
or anomaly-mediated SUSY breaking models. Of course, the extent to which
the ideas can actually be implemented in experiments depends on
how well these masses can be measured, and is thus sensitive to details of
the model.

The superpotential for the MSSM with a singlet neutrino superfield
$\hat{N}^c$ (for just a single generation), is given by
\begin{equation}
\hat{f}=\hat{f}_{MSSM}+f_{\nu}\epsilon_{ij}\hat{L}^i\hat{H}_u^j\hat{N}^c
+{1\over 2}M_N\hat{N}^c\hat{N}^c
\label{WmssmN}
\end{equation}
where $\hat{N}^c$ is the superfield whose fermionic component is the
left handed anti-neutrino and scalar component is $\tnu_R^\dagger$.
The soft SUSY breaking terms now include
\begin{equation}
{\cal L}={{\cal L}_{MSSM}}-m_{\tnu_R}^2 |\tnu_R |^2 +\left[
A_{\nu}f_{\nu}\epsilon_{ij}\tilde{L}^i\tilde{H}_u^j\tilde{\nu}_R^\dagger
+ \frac{1}{2} B_{\nu} M_N \tnu_R^2 + h.c. \right].
\label{LmssmN}
\end{equation}
The parameters $A_{\nu}$, $B_{\nu}$
and $m_{\tnu_R}$ are assumed to be comparable to the weak scale, even
though the right handed neutrino and its superpartner have a mass close
to $M_N$ which is much larger.

The complete set of 1-loop RG equations (RGEs) for the MSSM extended by a 
RHN may be found in
Ref.~\cite{jhep}, while the 2-loop ones are presented in Ref. \cite{imh2}.
In Ref. ~\cite{jhep}, sample sparticle mass spectra are shown with and
without the effect of a neutrino Yukawa coupling.
The 1-loop RGEs for slepton and left handed sneutrino masses contain
contributions from gauge interactions and from Yukawa interactions. The
former is generation independent while the latter is significant for
just the third generation. Furthermore, this Yukawa contribution
contains an additional term from the neutrino Yukawa coupling if the
model contains a RHN. We begin by defining the quantities
\begin{eqnarray*}
\Delta_R = m_{\te_R}^2-m_{\ttau_R}^2,\\
\Delta_L = m_{\te_L}^2-m_{\ttau_L}^2,
\end{eqnarray*}
where the $m_i$ denote the soft-SUSY breaking mass parameters. Aside
from $D$-terms, these are essentially the sparticle masses for the first
two generations. This is not the case for the third generation because
of effects of Yukawa interactions.  Including effects from the RHN, the
1-loop\footnote{We present the 1-loop RGEs to explain our strategy. In
our numerical analysis, however, 2-loop RGEs are used.} RGEs for these
quantities are given by\cite{jhep},
\begin{eqnarray}
d\Delta_R \over dt &=&{2\over{16\pi^2}}\Big(2 f_{\tau}^2 X_{\tau} \Big),
\label{eq:delR}\\
d\Delta_L \over dt &=&{2\over{16\pi^2}}\Big(f_{\tau}^2 X_{\tau}+
f_{\nu}^2 X_{\nu}\Big), \label{eq:delL}
\end{eqnarray}
where $t=\ln Q$, $X_{\tau}=m_{\ttau_L}^2+m_{\ttau_R}^2 +
m_{H_d}^2+A_{\tau}^2$ and $X_{\nu}=m_{\ttau_L}^2+m_{\tnu_R}^2 +
m_{H_u}^2+A_{\nu}^2$. 
Of course, below the scale $M_N$ the theory reduces to the MSSM, and the
last term in Eq. (\ref{eq:delL}) is absent. This then implies that in a
theory with a RHN, 
\begin{equation}
{d \over dt}(2\Delta_L-\Delta_R)={4\over 16\pi^2}f_{\nu}^2 X_{\nu},
\label{main}
\end{equation}
for $M_{GUT} \ge Q \ge M_N$.

Eq.~(\ref{main}) is the starting point for
our analysis. Within mSUGRA, $2\Delta_L-\Delta_R \simeq 0$ since the
masses are degenerate at the GUT scale and this quantity does not
evolve at 1-loop.
In contrast, for a model with a large
RHN Yukawa coupling, this quantity could evolve significantly between
$M_{GUT}$ and $M_N$. The value of $2\Delta_L-\Delta_R$ at the weak scale
is, of course, its value at $Q=M_N$, since it does not evolve for $Q
\leq M_N$. 

This is illustrated in Fig.~\ref{fig:delLR}, where we show
\begin{equation}
\delta_{LR} \equiv  {{2\Delta_L -\Delta_R} \over
{(m_{\te_L}^2+m_{\te_R}^2+m_{\ttau_L}^2+m_{\ttau_R}^2)/4}},
\label{delLR}
\end{equation}
which is just the dimensionless rendition of $2\Delta_L -\Delta_R$
obtained by dividing by the mean squared mass of the relevant four
quantities, in ({\it a})~the mSUGRA model, and ({\it b})~the model with
a RHN. In frame ({\it a}) we show $\delta_{LR}$ for 2400 randomly
generated mSUGRA models for the parameter ranges,
\begin{displaymath}
10~{\rm{GeV}} \leq m_0 \leq 1500~{\rm{GeV}},\\
10~\rm {GeV} \leq m_{1/2} \leq 500~\rm{GeV},\\ 
-1.8m_0 \leq A_0 \leq 1.8m_0,\\
\mu = +,-,
\end{displaymath}
while in frame ({\it b}) we generate models that contain an additional
RHN.  For definiteness, we assume that the additional soft SUSY breaking
parameters are unified at the GUT scale, and further, that
$f_{\nu}(M_{GUT})=f_t(M_{GUT})$. Then, the only free parameter is the
superpotential mass $M_N$ for the singlet neutrino, which we
vary\footnote{If we assume a simple see-saw mechanism for neutrino
masses, $M_N$ is constrained by Super-Kamiokande data. We will return to
this issue later.} between $10^5$~GeV and $M_{GUT} = 2\times 10^{16}$~GeV. For
every set of model parameters, and for both models, we compute
$\delta_{LR}$ and plot it against $M_N$ in Fig.~\ref{fig:delLR}.  Of
course, there is no RHN in mSUGRA; in the plot in frame ({\it a}), each
parameter set is randomly assigned a value of $M_N$, resulting in the
$M_N$ independent distribution of points in this case. The important
thing to note is that $\del_{LR}$ is negative and typically smaller in
magnitude than 0.01 in frame ({\it a}), and essentially always smaller
than 0.02. In contrast, in models with a RHN, $\delta_{LR}
\sim$~0.1-0.6, except when $M_N$ is very close to $M_{GUT}$ in which
case there is no range for RG evolution to occur.
This is in keeping with expectations from a single step
integration of Eq.~(\ref{main}) which gives,
\begin{displaymath}
(2\Delta_L-\Delta_R)_{M_N} \approx {{4}\over {16\pi^2}}f_{\nu}^2 X_{\nu}
\ln {{M_{GUT}}\over {M_N}},
\end{displaymath}
assuming all SUSY breaking parameters
with dimensions of mass are comparable.

\subsection{Effects of Stau Mixing}
While the idea described above is conceptually simple, the problem as we
have already noted is that SM Yukawa interactions cause
intra-generational mixing and preclude direct determination of
$m_{\ttau_L}$ and $m_{\ttau_R}$, and hence of $\delta_{LR}$.  Motivated
by the fact that the lighter (heavier) tau slepton is mainly $\ttau_R$
($\ttau_L$) we construct new variables,
\begin{eqnarray*}
\Delta_1 &=&m_{\te_R}^2-m_{\ttau_1}^2, \\
\Delta_2 &=& m_{\te_L}^2-m_{\ttau_2}^2,\\
\Delta_{\nu}& =&   m_{\tnu_{e}}^2-m_{\tnu_{\tau}}^2,
\end{eqnarray*}
that can, in principle, be directly measured in experiments at future
linear colliders. Since $\tnu_L-\tnu_R$ mixing is negligible,
$\Delta_{\nu}$ defined above is equal to $\Delta_L$, and we have only
introduced it for notational reasons.  Unlike the case of soft masses
where $SU(2)$ symmetry implied $m_{\ttau_L} = m_{\tnu_{\ttau}}$, we now
have three logically independent ``flavour differences'' that can be
formed (see Eq.~(\ref{defn}) below). Notice that the hypercharge
$D$-term contribution to the masses always cancels as it is
generation-independent. We can now form dimensionless differences
$\delta_{12}$, $\delta_{1\nu}$, and $\delta_{12\nu}$ analogous to
$\delta_{LR}$ defined above: 
\begin{eqnarray}
\delta_{12} &=& {{2\Delta_2 -\Delta_1}\over
{(m_{\te_L}^2+m_{\ttau_2}^2+m_{\te_R}^2+m_{\ttau_1}^2)/4}}, \nonumber \\
\delta_{1\nu} &=& {{2\Delta_{\nu} -\Delta_1}\over
{(m_{\tnu_e}^2+m_{\tnu_{\tau}}^2+m_{\te_R}^2+m_{\ttau_1}^2)/4}}, \\
\delta_{12\nu} &=& {{\Delta_2+\Delta_{\nu} -\Delta_1}\over
{(m_{\te_L}^2+m_{\ttau_2}^2+m_{\tnu_e}^2+m_{\tnu_{\tau}}^2+m_{\te_R}^2
+m_{\ttau_1}^2)/6}}, \nonumber 
\end{eqnarray}
which, we emphasize, might be directly measurable. In the
limit that stau mixing is negligible, the analysis
should essentially reduce to that which was previously considered. The
issue then is to examine how much stau mixing changes the variables
$\delta_{12}$, $\delta_{1\nu}$ and $\delta_{12\nu}$ from their values
(close to zero in mSUGRA) in the absence of mixing.

Toward this end, we show each of these quantities in Fig.~\ref{fig:mix}
for the mSUGRA model in the first column, and for the mSUGRA model
extended by a RHN in the second column. The plot is made in the same way as
Fig.~\ref{fig:delLR}. The following features are worthy of note.
\begin{itemize}
\item The spread of the points in the mSUGRA model is considerably
larger than in Fig.~\ref{fig:delLR}. Furthermore, as can most easily be
seen from the out-liers, it is largest for $\delta_{12}$ and smallest
for $\delta_{1\nu}$, and is always negative.

\item In contrast to the corresponding situation in
Fig.~\ref{fig:delLR}, we see that mixing allows some of the $\delta$s
(especially $\delta_{12}$) to become negative, and the very clean
separation between the two models that we had in the previous figure no
longer obtains. From the point of view of distinguishing the framework with
a RHN from the mSUGRA framework, it is clear that $\delta_{1\nu}$ is the
most effective variable.
\end{itemize}

To understand why this is the case, we first write,
\begin{eqnarray}
2\Delta_2-\Delta_1 &=&
2\Delta_L-\Delta_R+2(m_{\ttau_L}^2-m_{\ttau_2}^2)-
(m_{\ttau_R}^2-m_{\ttau_1}^2), \nonumber \\
2\Delta_{\nu}-\Delta_1 &=&
2\Delta_L-\Delta_R-(m_{\ttau_R}^2-m_{\ttau_1}^2), \label{defn}\\
\Delta_2+\Delta_{\nu}-\Delta_1 &=&
2\Delta_L-\Delta_R+(m_{\ttau_L}^2-m_{\ttau_2}^2)-
(m_{\ttau_R}^2-m_{\ttau_1}^2), \nonumber
\end{eqnarray}
and then note that $m_{\ttau_L}^2-m_{\ttau_2}^2
=m_{\ttau_1}^2-m_{\ttau_R}^2$ (since the sum of mass squared eigenvalues
must be the trace of the stau mass squared matrix), and further, that each
of these terms is negative because mixing always decreases (increases)
the lowest (highest) eigenvalue. We thus expect that $\ttau_L-\ttau_R$
mixing effects reduce $\delta_{12}, \delta_{1\nu}$ and $\delta_{12\nu}$
in Fig.~\ref{fig:mix} relative to $\delta_{LR}$ in Fig.~\ref{fig:delLR},
and further, that this reduction is in the ratio 3:1:2. Since these
quantities are all negative within the mSUGRA framework, $\delta_{1\nu}$
in the second row of Fig.~\ref{fig:mix} provides the cleanest
separation. This then requires precise measurements of $m_{\te_R},
m_{\tnu_e}, m_{\ttau_1}$ and $m_{\tnu_{\tau}}$, the prospects for which
we will discuss in the next section.  
It is worth mentioning that stau
mixing effects discussed above are absent in the quantity
$3\Delta_{\nu}-\Delta_1-\Delta_2$ but its determination requires
knowledge of $m_{\te_L}$ and $m_{\ttau_2}$ as well.

Up to now, we have ignored the Super-Kamiokande measurement
$10^{-3}$~eV$^2 \leq \Delta m^2 \leq 10^{-2}$~eV$^2$ which, assuming an
inter-generational hierarchy of neutrino masses, implies
\begin{equation}
0.033 \ \rm{eV} \leq m_{\nu_{\tau}} \leq 0.1 \ \rm{eV}. 
\label{superK}
\end{equation}
Within the simplest see-saw model the neutrino mass is given by
$m_{\tnu} = (f_{\nu}v_u)^2/M_N$, where $v_u$ is the vacuum
expectation value of the field $h_u$ responsible for the masses of
up-type fermions in the MSSM. This in turn implies that $f_{\nu}$ and
$M_N$ are strongly correlated. In fact, if $f_{\nu}(M_{GUT}) =
f_t(M_{GUT})$ then $m_{\nu} \sim m_t^2/M_N$, so that values of $M_N
\alt 10^{14}$~GeV are excluded by the Super-Kamiokande data. Then, from
Fig.~\ref{fig:mix}, we see that $\delta_{1\nu}$ has to be determined to
within 0.05-0.1 in order to distinguish the RHN model from
mSUGRA.

We can, however, repeat the previous analysis ignoring the GUT
constraint on Yukawa couplings and treating $f_{\nu}$ and $M_N$ as
parameters of the RHN model, but instead constrain these to yield
$m_{\nu}$ in the range (\ref{superK}). As before, we generate random
parameter sets for both these models, and then compute $\delta_{12}$,
$\delta_{1\nu}$ and $\delta_{12\nu}$ introduced earlier. The results are
shown in Fig.~\ref{fig:neut}. Except for the more limited range of $M_N$
shown here, the three figures in the first column are essentially the
same as in Fig.~\ref{fig:mix}, as should be the case since changing the
assumptions about the RHN sector do not affect the mSUGRA study. The
results in the second column are, however, qualitatively different: the
largest difference between the models now occurs when $M_N$ is close to
$M_{GUT}$. This is not difficult to understand. Since we are holding
$f_{\nu}^2/M_N$ (roughly) constant, $f_{\nu}$ is largest when $M_N$ is
large, and decreases roughly as $\sqrt{M_N}$. If $M_N \alt
{\cal{O}}(10^{13})$~GeV, the neutrino Yukawa coupling becomes too small
to have a visible effect on the sparticle masses. However, for larger
values of $M_N$, $\delta_{LR} \propto f_{\nu}^2 \ln{{M_{GUT}}\over{M_N}}
\propto {M_N \ln{{M_{GUT}}\over{M_N}}}$ grows very rapidly\footnote{Of
course, if $M_N=M_{GUT}$ all the $\delta$s will vanish (except for
mixing effects) because there is
no RG evolution. The dots in
Fig.~\ref{fig:neut} terminate well before this because as $M_N$
increases, the value of $f_{\nu}$ becomes so large that it blows up
before $M_{GUT}$. In this case, the model is not shown.} as long as
$M_N$ is not too close to $M_{GUT}$. Stau mixing then reduces the $\delta$s
shown in Fig.~\ref{fig:neut} exactly as discussed for the last figure.
Once again, $\delta_{1\nu}$ offers the best separation between mSUGRA
and models with a RHN, and if $M_N$ is within a factor $\sim 100$ of
$M_{GUT}$ distinction between the models may be possible if sparticle
masses can be measured with sufficient precision.

We find it exciting that, respecting the mass constraint, models with
values of $M_N$ close to $M_{GUT}$ (which theoretical prejudice might
suggest are the most likely) are the ones most likely to yield to
experimental scrutiny. It is thus interesting to ask whether a subset of
these models also satisfies the Yukawa unification condition, and if so,
examine whether this ``favored class'' of models can be distinguished
from mSUGRA. This is illustrated in Fig.~\ref{fig:nugut} where we show
$\delta_{1\nu}$ for ({\it a})~mSUGRA, ({\it b})~mSUGRA + RHN with
$f_{\nu}(M_{GUT})=f_t(M_{GUT})$, ({\it c})~mSUGRA + RHN with the
neutrino mass constraint, and ({\it d})~mSUGRA + RHN with
$f_{\nu}(M_{GUT})=f_t(M_{GUT})$ and the neutrino mass constraint. The
first three frames are the same as in previous figures and are only
included for convenience. We see from frame ({\it d}) that it is indeed
possible to find models where the neutrino and top Yukawa couplings
unify (an $SO(10)$-like condition) and the tau neutrino mass is in the
right range. Furthermore, the bulk of these models have large
values~\footnote{We should mention that we made a dedicated run where we
generated models with sleptons lighter than 500~GeV. The results were
qualitatively similar to those in frame {\it d}, except that the 
region of highest density
was shifted down slightly. In particular none of the models in this new
run yielded $\delta_{1\nu} > 0.2$.} of $\delta_{1\nu}$ so that it is
possible that linear collider experiments may offer a novel check of the
simplest see-saw mechanism, if the relevant
sparticle masses can be measured precisely enough.

Before closing this section, we briefly remind the reader that we had
noted that stau mixing effects which are responsible for the few points in
Fig.~\ref{fig:nugut}{\it d} overlapping with the mSUGRA region would 
be absent for the variable $3\Delta_{\nu}-\Delta_1-\Delta_2$. Thus, the
dimensionless version of this, 
\begin{displaymath}
\delta'_{12\nu}={{3\Delta_{\nu}-\Delta_1-\Delta_2}\over{(m_{\te_R}^2+
m_{\te_L}^2 + m_{\tnu_e}^2+m_{\ttau_1}^2+m_{\ttau_2}^2+m_{\tnu_{\tau}^2})/6}},
\end{displaymath}
in principle offers the hope of an even cleaner separation between the
two classes of models. This is illustrated in Fig.~\ref{fig:clean} where
we plot $\delta'_{12\nu}$ for ({\it a})~mSUGRA, and ({\it b})~mSUGRA +
RHN model with both $f_{\nu}(M_{GUT})=f_t(M_{GUT})$ and the neutrino
mass constraint (\ref{superK}). We see a dramatic reduction in the
spread for mSUGRA models in frame ({\it a}) which then resembles the
first frame of Fig.~\ref{fig:delLR}, confirming that the spread in the
other figures indeed originates from the mixing. Frame ({\it b}) shows
that none of the RHN models that we generated gives $\delta'_{12\nu}<
0$, so that the separation is theoretically very clean. Of course, this
measurement also requires a determination of $m_{\ttau_2}$ and
$m_{\te_L}$ in addition to the other sparticle masses.

\section{Prospects for Third Generation Slepton Mass Measurements}

We have seen that a determination of $\delta_{1\nu}$ or, even better
$\delta'_{12\nu}$, offers an opportunity for detecting Yukawa
interactions of neutrinos, and further, for an {\it independent} 
confirmation of the simple see-saw mechanism for neutrino masses.
In the simplest case where the neutrino and top Yukawa couplings unify
at the GUT scale these quantities,
which are zero in mSUGRA, are expected to be $\sim 0.05-0.2$, depending
on the RHN model parameters.

Assuming that sparticles directly decay to the lightest SUSY particle
(LSP) which was taken to be the lightest neutralino $\tz_1$, it has been
shown\cite{murayama} that masses of the first two generations of charged
sleptons and charginos can be measured at the 1-1.5\% level if these
sparticles are within the kinematic reach of $e^+e^-$ linear
colliders. An integrated luminosity of about 50~$fb^{-1}$ is sufficient
to attain this precision.  Longitudinal polarization of the $e^-$ beam,
which is crucial both to reduce SM backgrounds as well as to separate
various SUSY processes from one another, is expected in all the current
machine designs. Subsequently, it was shown\cite{munroe} that this
precision was not degraded even if sparticles decay via the complicated
cascades\cite{bbkmt} expected in most SUSY models. Indeed cascade decays
are an asset in that they make it possible to determine, for instance,
$m_{\tnu_e}$ with a similar precision when $\tnu_e \to \tw_1 e$ is
kinematically allowed\cite{munroe}. These findings have since been
confirmed by  more
realistic studies\cite{snowmass,zdr}. Thus, whether or not $\delta_{1\nu}$ or
$\delta'_{12\nu}$ can be conclusively determined to be non-zero depends
mainly on the precision with which the masses of third generation
charged sleptons and sneutrinos can be measured. A qualitative
assessment of this forms the subject of this section.

Nojiri {\it et al.} \cite{mihoko} performed a detailed study of how well
$m_{\ttau_1}$ can be measured at a linear collider, assuming that
$\ttau_1 \to \tau\tz_1$. Just as for the first two
generations\cite{murayama,munroe}, their basic strategy for determining
$m_{\ttau_1}$ is to study the energy distribution of (hadronically
decaying) daughter $\tau$s. This study is complicated by the fact that
$\tau$ is unstable and part of its energy is carried off by the
undetected neutrino. As a result, the relatively sharp end-points that
were obtained in earlier studies of $\te_R$ and $\tmu_R$ masses are
washed out. Nevertheless, assuming an integrated luminosity
of 100~$fb^{-1}$, they are able to obtain a $1\sigma$ precision of $\sim
2$\% on $m_{\ttau_1}$. In their study, they confined themselves to the
decay $\tau \to \rho\nu$, and estimated that by including decays to
$\pi$ and $a_1$, they might be able to improve this by a factor of 2. 

We are not aware of any studies that examine prospects for measuring
$m_{\ttau_2}$ or $m_{\tnu_{\tau}}$. A systematic evaluation of this would
require a dedicated study in itself, and is beyond the scope of this
paper. However, in order to get some feel for whether the measurements
outlined in the last section are feasible, we attempted to make a rough
estimate of how well $m_{\tnu}$ might be measured for a ``typical''
mSUGRA case where visible decays $\tnu \to \tw_1\ell$ of the sneutrino
are allowed. We do not
compute SM physics backgrounds (which are argued to be small) and ignore
any QCD jets that may be misidentified as hadronically decaying taus. We
perform only one case study and ignore bremsstrahlung and beamsstrahlung. 
We also assume that $m_{\tw_1}$ and $m_{\tz_1}$ will be well
measured by the time there is a large enough data sample to make it
possible to consider measurements of $m_{\tnu_{\tau}}$ or $m_{\ttau_2}$,
and ignore any error in these masses. 
Our purpose in
showing these simplistic results is two-fold. First, as we said, we
wanted to get a feel for whether measurements of $\delta_{1\nu}$ or
$\delta'_{12\nu}$ are even feasible. Second, we found that naive
extensions of methods previously used\cite{munroe} for measurements of
masses of heavier sleptons of the first two generations do not seem to
work in this case. We felt it would be worthwhile to point this out and
suggest a possible alternative.

The parameter point that we choose corresponds to an mSUGRA model with
\begin{displaymath}
m_0=150~\rm{GeV}, m_{1/2}=170~\rm{GeV}, A_0=0, \tan\beta=5, \mu>0.
\end{displaymath}
The resulting spectrum is illustrated in Table \ref{spectrum}. We see
that at a $\sqrt{s}=500$~GeV $e^+e^-$ collider, all charged sleptons and
sneutrinos, as well as charginos and neutralinos together with all but
the charged Higgs bosons are kinematically accessible via $2 \to 2$
processes. 

The production cross sections for the most important SUSY processes
obtained using ISAJET \cite{isajet} are shown in
Fig.~\ref{fig:csections} as a function of the electron beam polarization
parameter $P_L = f_L-f_R$, where $f_L$ ($f_R$) is the fraction of left
handed (right handed) electrons in the beam. The first item of note
is that cross sections for third generation sfermions are
small, and moreover, their signals are likely to suffer from
contamination from production of other sparticles. Here, we begin by
considering the measurement of $m_{\ttau_1}$. Our purpose is not to
improve upon the results of the detailed study of Nojiri {\it et
al.}\cite{mihoko}, but to see how our simplified analysis compares with
their results in order to be able to assess our results for the
corresponding study for $\tnu_{\tau}$.

We use ISAJET v7.51 for our SUSY event simulation. 
We use a toy calorimeter covering
$-4<\eta <4$ with cell size $\Delta\eta\times\Delta\phi =0.05\times 0.05$.
Energy resolution for electrons, hadrons and muons is taken to be
$\Delta E=\sqrt{.0225E+(.01E)^2}$, $\Delta E=\sqrt{.16E+(.03E)^2}$ and
$\Delta p_T =5\times 10^{-4} p_T^2$, respectively.
Jets are found using fixed cones of size
$R=\sqrt{\Delta\eta^2+\Delta\phi^2} =0.6$ using the ISAJET routine
GETJET (modified for clustering on energy rather than transverse energy).
Clusters with
$E>5$ GeV and $|\eta ({\rm jet})|<2.5$ are labeled as jets.
Muons and electrons are classified as isolated if they have $E>5$ GeV,
$|\eta (\ell )|<2.5$, and the visible activity within a cone of $R=0.5$
about the lepton direction is less than
$max({E_{\ell}\over 10},\ 1\ {\rm GeV})$. 
Jets with $E \geq 10$~GeV and one or three charged tracks (with
$p_T(track) \ge 0.5$~GeV) in a 10$^\circ$ cone about the jet axis, but
no other tracks in a larger 30$^\circ$ cone, are classified as $\tau$s
if the mass of the tracks is smaller than $m_{\tau}$ and the total
charge of the tracks is $\pm 1$.

\subsection{$\ttau_1$  mass measurement}

In order to obtain a measurement of $m_{\ttau_1}$ it is clear from
Fig.~\ref{fig:csections} that it would be best to use $P_L(e^-)$ as
close to $-1$ as possible. Then potential SUSY
contamination to the $\tau\tau+\eslt$ final state\footnote{A $\tau$
final state always refers to the visible debris from a hadronically
decaying $\tau$.} from $\tw_1\tw_1$ and
$\tz_2\tz_2$ production, as well as from SM $W^+W^-$ production, is
minimized, resulting in a relatively clean sample of $\ttau_1\ttau_1$
events. We use $P_L(e^-)=-0.9$ corresponding to a 95\% polarized beam.
To selectively enhance the signal over SM backgrounds, we impose similar
cuts used in earlier studies\cite{murayama,munroe} of the $\tmu_R$
signal. The difference is that instead of the dimuon final state from
$\tmu_R\tmu_R$ production, we now have a di-tau final state in the ``two
narrow jets'' channel. We require, (i)~$E_{\tau}^{vis} < 200$~GeV,
(ii)~$E_{T\tau}^{vis}\geq 15$~GeV, (iii)~20~GeV $\leq E(visible) \leq
400$~GeV, (iv)~$|\cos\theta_{\tau}|\leq 0.9$,
(v)~$-Q_{\tau}\cos\theta_{\tau} \leq 0.75$, (vi)~$\theta_{acop} \leq
30^\circ$, and (vii)~$\eslt \geq 25$~GeV. We veto events with additional
jets. Here, $E_{\tau}^{vis}$ and $E_{T\tau}^{vis}$ refer to the energy and
transverse momentum of the hadronically decaying tau jet; {\it i.e.}
the visible portion of the decay products of the tau. 
Cut (v) greatly reduces backgrounds from $W^+W^-$ and also $e\nu W$
production. Unlike in the
earlier study of smuon production, where an additional cut
$|m_{\mu\mu}-M_Z|> 10$~GeV was imposed to eliminate backgrounds from
$ZZ$, $\nu\nu Z$ and $e^+e^-Z$ production, we have not imposed a mass
cut on the di-tau system. 

We have checked that $ZZ$ and $WW$ events yield a background of 0.3~$fb$
and 0.1~$fb$, respectively (to be compared with the signal of $\sim
8$~$fb$) and ignored $\nu\nu Z$ and $e^+e^-Z$ backgrounds. Other SM
physics sources of $\tau\tau+\eslt$ events include $Z^*\to\tau\tau$
production, and $e^+e^- \to Z(\to\nu\nu)+ h_{SM}(\to\tau\tau)$
production. The former, which (for hadronically decaying taus) has a
total cross section of $\sim 200$~$fb$ (before any cuts) is reduced to a
negligible level after our cuts (especially (vi) and (vii)). Our
simulation of this yields a cross section\footnote{While this background
may well be considerably larger than $4\times
10^{-3}$~$fb$ after radiation is included, we
expect that it is still negligible.} $\sigma \alt 4\times
10^{-3}$~$fb$. Although we have not simulated the latter, we have run
all SUSY and Higgs (including $Zh$) contributions for the case at hand
through our cuts. We find that $\ttau_1\ttau_1$ production contributes
90\% of the signal, and of the remaining 10\%, only about 15\% comes
from $Zh$ production.\footnote{About 60\% of the SUSY and Higgs
background comes from chargino and neutralino production (of this about
2/3 is from $\tz_1\tz_2$ production), and a quarter from sneutrino and
stau production, with $Zh$ making up the remainder.}  Since the lightest
Higgs boson of the MSSM is SM-like, 
we infer that the $Zh_{SM}$ background is small. Presumably,
the SUSY background can be further reduced by dedicated cuts, {\it e.g.}
on $m_{\tau\tau}$ which has to be smaller than $m_{\tz_2}-m_{\tz_1}$ for
di-taus from $\tz_2$ decays. In the following discussion of the
precision with which $m_{\ttau_1}$ can be measured, we ignore all SUSY
and SM backgrounds.

To determine $m_{\ttau_1}$, we study the visible energy spectrum of the $\tau$ jets
whose distribution is shown in Fig.~\ref{fig:tau1}{\it a} for the mSUGRA case
under study. If the entire energy of the tau (produced via the two body
decay of a scalar) could be measured, we would expect this spectrum to
be flat with sharp end points in a perfect detector -- the decay
kinematics then determines the end points in terms of $m_{\ttau_1}$ and
$m_{\tz_1}$ along with the beam energy. The energy spectrum in
Fig.~\ref{fig:tau1}{\it a} is far from flat because of the energy lost to
neutrinos. Since higher energy taus, on an average, lose greater energy
to neutrinos, the upper end point is greatly smeared. The loss of energy
to neutrinos also means that there will be signal events below the lower
end point. Thus the kinematic end points do not play a role in a
measurement of sparticle masses. Nevertheless, the shape and
normalization of the $E_{\tau}$ distribution is sensitive to
$m_{\ttau_1}$. 

To ascertain how well $m_{\ttau_1}$ can be determined, we generated
``theory'' data samples (each with about 80 times as many stau pairs as
expected for the signal case for an integrated luminosity of
100~$fb^{-1}$) for several values of $\ttau_1$ mass to determine the
theoretical $E_{\tau}^{vis}$ distribution after the cuts (the solid
histogram). We also generated a synthetic ``data'' set for an integrated
luminosity of 100~$fb^{-1}$ (the points with error bars). Assuming that
$m_{\tz_1}$ is well determined (via a fit to the much larger sample of
$\te_R\te_R$ and $\tw_1\tw_1$ pair events\cite{murayama}) we can now fit
the $E_{\tau}^{vis}$ distribution from the ``data'' to the
``theory'' in terms of a single parameter $m_{\ttau_1}$. The resulting
$\chi^2$ distribution is shown in Fig.~\ref{fig:tau1}{\it b}. The triangles
denote the actual value of $\chi^2$ that we obtained for each theory
point that we compared with. The solid curve is the best fit parabola to
all these triangles. The scatter indicates the error on the theory
which, as we described, was also obtained via a Monte Carlo
calculation. From the figure, we obtain the fitted value of the stau
mass to be $168.1\pm 2.6$~GeV (1$\sigma$). In view of the scatter of our
``theory'', we will be more conservative and work with the 90\% CL error
of $\pm 3.5$~GeV. Our results, are not incompatible with those of Nojiri
{\it et al.} The slightly smaller error that we get may be attributed to
the fact that we have assumed that the LSP mass is known\footnote{This
may not seem a good assumption because the error on the stau mass is not
much smaller than the expected error of $\sim 1$\% on $m(LSP)$. We will
see shortly that a much larger integrated luminosity is needed for any
measurement of the mass of $\tnu_{\tau}$. In this era, both the LSP and
$\ttau_1$ masses will be much better determined.} so that we could
perform a single parameter fit, and possibly also because only $\tau \to
\rho$ decays were used in their analysis.

\subsection{$\tnu_{\tau}$ mass measurement}

The analysis of Ref.\cite{munroe} suggests that it should
be possible to measure $m_{\tnu_{\tau}}$ in the same way that the
electron sneutrino mass was measured. The strategy for this measurement
was to focus on the sample of very clean events from the process,
$e^+e^- \to \tnu_e+\tnu_e \to e\tw_1 + e\tw_1 \to e\mu\nu\tz_1 + eqq\tz_1$, 
where
one of the charginos decays hadronically and the other leptonically. In
this analysis, the electron beam was taken to be left handed.
There was very little SM background or SUSY contamination to this
event sample, and a two parameter fit to the flat electron energy
distribution yielded $m_{\tnu_e}$ and $m_{\tw_1}$ with a precision of
just over 1\% (1$\sigma$). 

The same strategy suggests that we should focus on $\tau\tau\ell
jj+\eslt$ events from $\tnu_{\tau}\tnu_{\tau}$ production, where
$\ell=e,\mu$, and study the (visible) $\tau$ energy distribution in this
data sample (but, of course, change the beam polarization to be right
handed to reduce backgrounds). We attempted to do so (again assuming the
chargino mass will be well measured), but were unable to obtain
significant discrimination between data sets with different
$\tnu_{\tau}$ masses. We found several factors which cause a difference
from the study in Ref.\cite{munroe}.

\begin{enumerate} 
\item First, $\sigma(\tnu_e\tnu_e)$ is much larger than
$\sigma(\tnu_{\tau}\tnu_{\tau})$: for our case study we see from
Fig.~\ref{fig:csections} that this factor is larger than  100.

\item While the branching ratio for $\tnu \to \ell\tw_1$ and $\tw_1 \to
\ell\nu\tz_1$ are similar (in fact the latter favours our case as we can use both
$e$ and $\mu$ decays of $\tw_1$), we now have to require that both taus
decay hadronically (a reduction to $4/9$), and further, that both the tau jets
pass the identification criteria to be identified as taus.

\item Not all the energy of the tau is visible. In particular, since the
visible energy of the tau is reduced due to the escaping neutrinos, we
found that the energy spectrum is frequently pushed well below our tau
identification threshold of 10~GeV.

\item For a left handed electron beam $\sigma(\tnu_e\tnu_e)$ is much
larger than other SUSY cross sections, so that the event sample is very
much dominated by the sneutrino signal. This is not quite the case for
$\tnu_{\tau}$ pair production; we see from Fig.~\ref{fig:csections} that
heavier chargino and neutralino production as well as $\ttau_2\ttau_2$
production may make significant contributions to this event topology.

\end{enumerate}
The first two items greatly reduce the number of signal events
increasing the statistical error. The last item causes contamination of
the sample. The energy loss due to the escaping neutrinos discussed in
item (3) causes the energy distributions with different
$m_{\tnu_{\tau}}$ to resemble one another beyond $E_{\tau}=10$~GeV more
closely than they would if the full $\tau$ energy could be measured. The
point is that the cut on the tau energy cuts out fewer events when the
mass gap between $\tnu_{\tau}$ and $\tw_1$ increases; {\it i.e.} for
heavier sneutrinos, assuming as we do that the chargino mass is fixed.
But the sneutrino production cross section reduces for larger sneutrino
mass. As a result, these two effects compensate, reducing the
discrimination between different sneutrino mass cases.\footnote{Of
course, at the high energy end the spectra are sensitive to the
sneutrino mass, but there we do not have sufficient event rate to
significantly contribute to $\chi^2$ in an analysis similar to the one
for $m_{\ttau_1}$ determination discussed above.} We also examined other
decay chains (with just one identified $\tau$) but found that these
generally suffered from large contamination from other SUSY sources. We
conclude that this strategy is not suitable for such a measurement even
with an integrated luminostiy of 500-1000~$fb^{-1}$.

The energy dependence of the cross section provides an alternative way
to measure the sneutrino mass. Since we have not taken bremsstrahlung or
beamsstrahlung
into account, we are careful not to go very close to the kinematic
threshold where the cross section would be most strongly affected by soft
photon emission. Away from the threshold, we expect that beamsstrahlung
effects may reduce the cross section by $\sim 10$\%, but presumably without
greatly altering its energy dependence. 

The total cross section for $\tau\tau\ell jj+\eslt$ events from all SUSY
sources is shown in Fig. \ref{fig:snuall}{\it a} for the signal point
with $m_{\tnu_{\tau}}=178.1$~GeV, as well as for two other values of the
sneutrino mass. Here, we have required that each $\tau$ jet has
$E_{\tau}^{vis} \ge 10$~GeV. We also require $\eslt \ge 25$~GeV. SM
physics backgrounds are expected to be small. The error bars correspond
to an integrated luminosity of 200~$fb^{-1}$ for each energy. For the
contamination from SUSY events, we have added contributions from other
SUSY sources but with mSUGRA parameters fixed to their case study
values\footnote{This is a conservative attitude, because if we compute
the SUSY background (mainly from heavier chargino and neutralino
production at the highest energies) using the different mSUGRA
parameters for each sneutrino mass case, the {\it SUSY background cross
sections} change enough to allow a discrimination between the
scenarios. This occurs because of the change in chargino and neutralino
masses, but these will be well determined by the time tau sneutrino mass
determination might become possible. Our assumption is, perhaps,
ultra-conservative in that effects from a change in the sneutrino mass,
which also change the branching fractions of charginos and neutralinos,
are ignored.}  regardless of $m_{\tnu_{\tau}}$.

The cross section for sneutrino events, along with that for the other
important contributors to this topology, is shown in Table~\ref{back} for
the energy range in Fig.~\ref{fig:snuall}. We see that the contribution from
$\ttau_2\ttau_2$ production is always much smaller than that from
charginos and neutralinos. For this case, the SUSY backgrounds are
smaller than or comparable to the signal for $\sqrt{s} \alt 550$~GeV,
but for larger values of $\sqrt{s}$ the heavy chargino and neutralino
contributions overwhelm the sneutrino signal. The sharp increase in the
cross section from charginos and neutralinos that is seen for $\sqrt{s}
\geq 550$~GeV is due to the opening up of heavy neutralino and chargino
pair production thresholds. 

To get an idea of how well cross sections can distinguish different
sneutrino masses, we perform a Monte Carlo computation of the cross
section for several values of sneutrino mass for energies ranging
between 425~GeV and 600~GeV in steps of 25~GeV. We generate about $10^5$
SUSY events which is about 30 times the number of signal events expected
for an integrated luminosity of 200~$fb^{-1}$. For each value of
$m_{\tnu_{\tau}}$ that we consider, we then compute $\Delta\chi^2$
between the energy dependence of the cross section for the case study
point and the corresponding quantity for some other $\tnu_{\tau}$
mass. Our results are shown in Fig.~\ref{fig:snuall}{\it b} for a
statistical error bar corresponding to an integrated luminosity of
100~$fb^{-1}$ per point (circles) and 200~$fb^{-1}$ per point
(triangles). Also shown is the line corresponding to $\Delta\chi^2=2.7$
(the 90\% CL in the Gaussian limit). The solid (dashed) curves are a fit
through the circles (triangles). Using these fits, we see that the
sneutrino mass is obtained as $m_{\tnu_{\ttau}} =178^{+15}_{-18}$~GeV (
$m_{\tnu_{\ttau}}=178^{+10}_{-13}$~GeV) for an integrated luminosity of
100~$fb^{-1}$ (200~$fb^{-1}$) at the 90\% CL.\footnote{The
$\Delta\chi^2$ distribution is not exactly symmetric, so that one should
not view the 90\% CL literally, but regard the result as a qualitative
indicator of the precision of the measurement.} We recognize that even
with an integrated luminosity of 100~$fb^{-1}$ per point, it would take
several years of running to obtain the 800~$fb^{-1}$ of integrated
luminosity that would be needed, assuming current projections for the
anticipated luminosity of such a machine. It is worth keeping in mind
that new developments ({\it e.g.} vertex detection, neural net
algorithms, or something else) will, quite possibly, result in a
significantly higher efficiency for tau identification than assumed in
our analysis. In this case, the integrated luminosity required will be
correspondingly reduced.  Our purpose, however, is not to imply that
$\tnu_{\tau}$ mass measurement is possible, but to clarify just how
challenging such a measurement might be.

The analysis above assumes that it would not be possible to sort out
chargino and neutralino events from $\tnu_{\tau}$ events. Since the
heavier chargino $\tw_2$ and the heavier neutralinos $\tz_{3,4}$
dominantly decay to real $W$, $Z$ or $h$ and light charginos and
neutralinos, it might be conceivable that by the time such large data
samples become available, experimentalists might have learnt to
recognize (the bulk of) $\tw_1$ and $\tz_2$, in the same way that they
recognize $b$-jets and $\tau$-leptons today. Alternatively, although the
case study point allows for heavy chargino and neutralino production, it
could be that for another point, SUSY contamination from these sources
is kinematically supppressed.
In either case, the chargino
and neutralino sample would then not contaminate the sneutrino sample. To see
how much this would improve the sneutrino mass measurement, we have
redone the analysis in Fig.~\ref{fig:snuall}, but assumed that there is
no background in the $\tau\tau\ell jj$ channel.\footnote{This is of
course overly optimistic because $\ttau_2\ttau_2$ events would be
kinematically similar to sneutrino events. Also, $\tz_4$ and $\tw_2$
have a branching fraction of $\sim 3$\% to decay into $\tnu_{\tau}$.}
Our results are shown in Fig.~\ref{fig:snuonly} where we have taken the
integrated luminosity to be 100~$fb^{-1}$ per point. As before, the
upper frame shows the cross section and the lower one $\Delta\chi^2$,
defined the same way as in Fig.~\ref{fig:snuall}. We see from
frame ({\it a}) that measurements at the lower energy can readily
distinguish very heavy sneutrinos, as their production is kinematically
suppressed. But these measurements do not give as sharp a distinction for
sneutrinos lighter than the test case.\footnote{ We could go to yet
lower energy, but we constrained ourselves to stay away from the
threshold. Of course, if backgrounds can be really eliminated, or
reliably subtracted, an energy scan close to the sneutrino threshold 
might yield the 
maximum precision.} For this case then, measurements
at the high energy end improve the discrimination. This is manifested in
Fig.~\ref{fig:snuonly}{\it b} where we show $\Delta\chi^2$ for a scan of
425-550~GeV for which the signal exceeds the SUSY background
(triangles), and also after including the highest two energy points (circles)
keeping in mind that the sneutrino sample is then highly contaminated.
Indeed, we see that the two curves match at the high mass end, but
including the 575 and 600~GeV bins significantly improves the
discrimination for lower values of $m_{\tnu}$. We see
that if SUSY backgrounds can be controlled, a sneutrino mass measurement
of comparable precision as in Fig.~\ref{fig:snuall} might be possible
with half the integrated luminosity.  

We conclude that a precise determination of $m_{\tnu_{\tau}}$ poses a
formidable challenge. Naively following the ideas in Ref. \cite{munroe} that
worked so well for a measurement of $m_{\tnu_e}$ simply does not
work. It appears possible that the energy dependence of the cross
section for $\tau\tau\ell jj$ $+\eslt$ events might allow a mass
measurement at the 8-10\% level, but such measurements would take
several years with current projections for the integrated
luminosity. With a data sample of 1600 $fb^{-1}$ distributed over 8
energy points a 6-7\% measurement appears possible. Alternatively, if
we can distinguish chargino/neutralino from sneutrino initiated events
without substantial loss of signal, a similar precision might be
possible with just half this luminosity. We should remember that we are
conservatively quoting 90\% CL and not 1$\sigma$ errors, partly because
of the simplified way that we have done our calculation.

We have not attempted to do an analysis of a measurement of
$m_{\ttau_2}$ which decays via $\ttau_2\to \tau\tz_1$ (10\%),
$\tau\tz_2$ (37\%) and $\nu \tw_1$ (53\%). The dominant decays lead to
$\tw_1\tw_1+\eslt$ events which have a large contamination from
$\tw_1\tw_1$ production. This could be reduced by using a right handed
electron beam, but even for 95\% polarization, the cross section from
direct chargino production exceeds that for chargino production via
heavy stau pair production a factor of about 3. Moreover, other
SUSY sources ($\tz_2\tz_2$, $\tw_1\tw_2$ and even
$\tnu_{\tau}\tnu_{\tau}$ production) could contaminate the signal
especially from hadronic decays of daughter charginos. While it is
possible that a clever use of the other decays modes of $\ttau_2$ may
well allow a measurement of its mass, we believe that this poses an even
greater challenge than the measurement of the $\tnu_{\tau}$ mass. 

\section{Can Slepton Masses Probe Right Handed Neutrino Couplings?}

We saw in Sec.~II that the variables $\delta_{1\nu}$ and
$\delta'_{12\nu}$ offer the best hope for measuring effects of the
sneutrino Yukawa coupling. While the latter is theoretically cleaner in
that models with and without a RHN are better separated, its determination
requires a measurement of $m_{\ttau_2}$ in addition to $\ttau_1$,
$\tnu_{\tau}$ and other slepton masses. Since we have just explained the
difficulties associated with this measurement, we will focus on the
prospects for the determination of just $\delta_{1\nu}$ in the remainder
of this paper. Within the simplest see-saw model with unified top and
$\nu_{\tau}$ Yukawa couplings at $M_{GUT}$, Fig.~\ref{fig:nugut}{\it d}
shows that a large fraction of RHN models have $\delta_{1\nu} \agt
0.1$. The question then is whether experiments at linear colliders will
have the sensitivity to distinguish $\delta_{1\nu}=0$ from
$\delta_{1\nu} \agt 0.1$.

It is straightforward to check\footnote{The variance of a function
$f(x_1,x_2,...x_n)$ of the uncorrelated quantities $x_i$ is $\sum
\left({{\partial f} \over {\partial x_i}}\right)^2 (\Delta x_i)^2$,
where the derivative is evaluated at the mean value of $x_i$, and
$(\Delta x_i)^2$ is the variance of $x_i$.} that if the error in
measuring first generation masses is negligible and mass differences
between the various sleptons are small, the error in
$\delta_{1\nu}$ is given by
\begin{equation}
(\Delta \delta_{1\nu})^2 = 16\left({{\Delta  m_{\nu_{\tau}}}\over
{ m_{\nu_{\tau}}}}\right)^2 + \left( {{\Delta m_{\ttau_1}} \over {m_{\ttau_1}}}
\right )^2. 
\label{error}
\end{equation}
Even neglecting the error in the measurement of $m_{\ttau_1}$, it
appears that $m_{\nu_{\tau}}$ needs to be determined at about the 2.5\%
level in order to obtain the required precision on
$\delta_{1\nu}$. Unfortunately, in view of the analysis in the previous
section this does not seem to be possible, at least using the techniques
considered in this study. Even if we change the 90\% CL error bar to the
``1$\sigma$'' error bar ($\Delta\chi=1$) the error on $m_{\tnu_{\tau}}$
seems to be about 4\%. It thus appears to us that the effects of the tau
neutrino Yukawa coupling on sparticle masses seems to be beyond the
projected sensitivity of linear collider experiments, at least within
the framework of the simplest GUT model with unification of neutrino and
top Yukawa couplings. However, if we allow the
neutrino Yukawa coupling at the GUT scale to be somewhat larger than
$f_t$ (but still require $f_{\nu}$ not to blow up before $M_{GUT}$) then
linear collider experiments could be sensitive to the presence of the
RHN as can be seen from Fig.~\ref{fig:neut}{\it b}.

In obtaining this conclusion, we assumed that the only information we
would have is on the masses of various sleptons. It could, however, be
that a measurement of $\tan\beta$ may also be 
possible\cite{murayama,barger}. 
In this case, we would not need to combine the masses to eliminate the
($\tan\beta$-dependent) effects of the usual tau Yukawa coupling so
that, in principle, a direct measurement of stau and $\tnu_{\tau}$
masses would contain information about $f_{\nu}$. We checked
that in all the models we generated that satisfied the Super-Kamiokande
neutrino mass constraint {\it and} had $f_t=f_{\nu}$ at the unification
scale, {\it i.e.} the models in frame ({\it d}) of Fig.~\ref{fig:nugut},
$m_{\tnu_{\tau}}$ and $m_{\ttau_2}$ differ from their mSUGRA values by
less than  7\% (typically 3-5\%) while $m_{\ttau_1}$ almost always differs
by less than 3\% (since $\ttau_1 \sim \ttau_R$, we expect it to be less
affected by $f_{\nu}$). We thus conclude that even in this case, it
would be very difficult to discern the effect of the neutrino Yukawa
coupling, except maybe for a small sub-set of model parameters.

This pessimistic outlook hinges upon the assumption of the simplest
see-saw model. It is, however, worth keeping in mind that there are
other models for neutrino masses that have been suggested. For instance,
in a model\cite{type3} with an additional $SO(10)$ singlet with mass
$M_S$ in the superpotential, and a GUT Higgs sector comprising of just
{\bf 16} dimensional representations, the light neutrino mass would take
the form $m_{\nu} = M_S m_t^2/M_N^2$, so that neutrino masses are still
hierarchichal but depending on $M_S$, $M_N$ would be considerably
smaller than in the case of the usual see-saw.\footnote{This mechanism
has been dubbed the type III see-saw, with the usual see-saw mechanism
being the type~I see-saw. In this context, we remark that for the
type~II see-saw \cite{mohapatra} we have the usual $2\times 2$ neutrino
mass matrix, but with an induced non-zero Majorana mass $m_L$ for the
left handed neutrino which contributes additively to the physical light
neutrino mass. This new contribution could be independent of the usual
Yukawa couplings, and so may be generation-independent, giving
(approximately) degenerate neutrinos of mass $m$ when $m \gg
m_{u,c}^2/M_N$. The splitting between neutrinos is still about
$m_t^2/M_N$ so that $\Delta m^2$ measured in neutrino oscillation
experiments is now $\sim 2m m_t^2/M_N$, which requires even larger $M_N$
than the usual see-saw to satisfy the Super-Kamiokande measurement. Thus
with the type~II see-saw, we do not expect to see measurable effects in
the slepton sector.}  In the extreme case\cite{DT} where $M_S \sim
1~TeV$, the RHN could be as light as $10^8$~GeV. In this case, we see
from Fig.~\ref{fig:nugut}{\it c} that $\delta_{1\nu}=0.2-0.5$ which is
in the range that experiments at linear colliders should be sensitive
to.

We also looked at the analysis\cite{neubert} of neutrino masses within
the framework of a localized gravity model with non-factorizable
geometry, where it was shown that neutrino masses were given by $m_{\nu_i}
\sim M \times ({{v}\over {M}})^{r_i+1/2}$. Here, the compactification
scale $M \sim M_{Planck}$, $v$ is the electroweak $vev$ and $r_i$ a real
number $\geq 1/2$. For $r_i=3/2$, this reduces to the familiar see-saw-like
formula. One might, at first glance, think that by adjusting $r_i$ it would be
possible to allow smaller values of $M$ which could then be probed via
slepton masses. This is not the case because in this framework lepton
number is conserved, and neutrinos only get a Dirac mass. In other
words, the induced neutrino Yukawa coupling is tiny, essentially because
of the small overlap between the active left handed neutrino (which is
confined to the brane) and the sterile neutrino in the bulk. We thus
expect slepton masses to be unchanged from their mSUGRA values in such a
scenario. 

In summary, we examined the effects of neutrino Yukawa couplings on the
masses of sleptons and sneutrinos present in supersymmetric models.  
For
most of the analysis, we assumed the simplest see-saw model for neutrino
masses, and also worked within the SUSY GUT framework which implies a
hierarchy of neutrino masses. Assuming that third generation neutrinos
are the heaviest,
we then expect the largest effect amongst third generation
sleptons. To separate the effect of the MSSM tau Yukawa coupling from
the new neutrino Yukawa coupling, we constructed several combinations of
masses, which are expected to be zero in the mSUGRA model but deviate
from this in the RHN framework.
Our results are shown in Fig.~\ref{fig:mix} as a function of
the RHN mass scale $M_N$, assuming that the top and neutrino Yukawa
couplings unify at the GUT scale.  The Super-Kamiokande atmospheric neutrino
data, interpreted as neutrino oscillations, however, implies that
$m_{\nu_{\tau}}$ is between 0.033 and 0.1~eV. In this case $M_N$ cannot
be too far from $M_{GUT}$. The best discrimination is obtained via the
variables $\delta_{1\nu}$ and $\delta'_{12\nu}$. Fig.~\ref{fig:nugut}{\it d}
and Fig.~\ref{fig:clean}{\it b} show that experiments should be
sensitive to the difference between 0 and about 0.06-0.1 in order to
conclusively discriminate RHN models from mSUGRA. Toward this end, we
did a simplified analysis of the precision with which third generation
sparticle masses might be measured at future $e^+e^-$ colliders. While
we confirmed the conclusions of previous analyses that $\ttau_1$ could
be measured with a precision of $\sim 2$\%, we found that with the
techniques that we examined $m_{\tnu_{\tau}}$ could at best be measured
with a precision of $\sim 6-8$\%, even with optimistic projections for
the luminosity and what might be achievable in the future. This then led
us to conclude that such mass measurements would not be able to
discriminate models with a RHN from mSUGRA within this simple
framework. We saw however, that if we give up the unification of top and
neutrino Yukawa couplings, or allow a more complicated framework (the
type~III see-saw) such a discrimination might be possible. 

We conclude that precision measurements of charged slepton and sneutrino
masses can provide information about RHN masses and couplings. For
instance, a non-vanishing value of $\delta_{1\nu}$ (or $\delta'_{12\nu}$) would
provide strong confirmation of large Yukawa interactions of neutrinos,
and hence an underlying GUT scale see-saw type mechanism. This would
then eliminate whole classes of alternatives including: ({\it i})
neutrino masses have Majorana type contributions only, ({\it ii})
neutrino masses are purely Dirac with Yukawa couplings strongly
suppressed for symmetry\cite{hall} or geometric\cite{neubert} reasons,
or ({\it iii})~there might be a TeV scale\footnote{Within this
framework, right handed
neutrinos and sneutrinos are at the TeV scale. The presence of a weak
scale $\tnu_R$ can result in very different sneutrino phenomenology
which might be probed at colliders, or via its cosmological
implications\cite{hall}.} see-saw, again with very suppressed
(effective) Yukawa interactions of neutrinos\cite{hall}: in all these
cases, we would expect that $\delta_{1\nu} = \delta_{12\nu} \simeq 0$.

While much attention has been focussed on measurements of charginos,
neutralinos, and first generation sparticle masses, there have been few
studies for the third generation of sfermions. On the
other hand, it is just these masses (not only sleptons and sneutrinos, but
also squarks) which are frequently sensitive to new physics at the high
scale (especially physics associated with family structure). This study
exemplifies the need to develop new techniques 
to measure third generation sparticle properties more precisely.

%
\acknowledgments We are grateful to M.~Nojiri, S. Pakvasa and M.~Peters
for discussions.  This research was supported in part by the
U.~S. Department of Energy under contract number DE-FG02-97ER41022 and
DE-FG-03-94ER40833.
%

%

%
\newpage
%
%

\iftightenlines\else\newpage\fi
\iftightenlines\global\firstfigfalse\fi
\def\dofig#1#2{\epsfxsize=#1\centerline{\epsfbox{#2}}}

\begin{table}
\begin{center}
\caption{Sparticle masses for the mSUGRA case study of Sec.~III. Squarks
are not accessible for the entire range of energies that we consider,
except for stop; $m_{\tst_1}=274.7$~GeV so that its pair production would
be accessible for $\sqrt{s} \geq 550$~GeV. Second generation slepton
masses are the same as those of the first generation.}
\bigskip
\begin{tabular}{cc|cc|cc}                 
particle &  $m$~(GeV)& particle & $m$~(GeV)& particle &  $m$~(GeV) \\         
\hline 
$\te_R$ & 167.8 & $\tz_1$ & 59.9 & $h$ & 105.0 \\
$\te_L$ & 194.2 & $\tz_2$ & 108.2 & $H$ &312.6 \\
$\tnu_e$ & 178.3 & $\tz_3$ & 255.5 & $A$ & 310.4\\
$\ttau_1$ & 165.7 & $\tz_4$ & 284.3 &$H^{\pm}$ & 320.4\\
$\ttau_2$ & 195.4 & $\tw_1$ & 105.6 & $\tu_L$ & 397.5\\
$\tnu_{\tau}$ & 178.1& $\tw_2$ & 283.4 & $\tg$ & 427.8
\label{spectrum}
\end{tabular}
\end{center}
\end{table}

\begin{table}
\begin{center}
\caption{Cross sections in $fb$ for the $\tau\tau\ell jj$ signal from
$\tnu_{\tau}\tnu_{\tau}$ production for the case study of Sec.~III, as a
function of the center of mass energy $\sqrt{s}$ after the cuts
described in the text. Also shown are the corresponding cross sections
from $\ttau_2\ttau_2$ production, and from chargino and neutralino
production. As discussed in the text, the cross sections in the last two
columns have been used as the background for {\it all} sneutrino masses
in Fig.~\ref{fig:snuall}.}
\begin{tabular}{c|c|c|c}
$\sqrt{s}$ (GeV) & $\sigma (\tilde{\nu}_{\tau} \tilde{\nu}_{\tau})$ (fb) &  
$\sigma (\tilde{\tau}_2 \tilde{\tau}_2)$ (fb)  & 
$\sigma (\tw_i \tw_j, \tz_i \tz_j)$ (fb)  \\
\hline
425  &  0.083  & 0.008  & 0.053  \\
450  &  0.116  & 0.015  & 0.071  \\
475  &  0.140  & 0.019  & 0.078  \\
500  &  0.139  & 0.028  & 0.079  \\
525 &   0.149  & 0.029  & 0.078  \\
550  &  0.149  & 0.036  & 0.127  \\
575  &  0.145  & 0.032  & 0.219  \\ 
600  &  0.134  & 0.037  & 0.270  
\label{back}
\end{tabular}
\end{center}

\end{table}

\newpage
%

%
\begin{figure}
\dofig{6in}{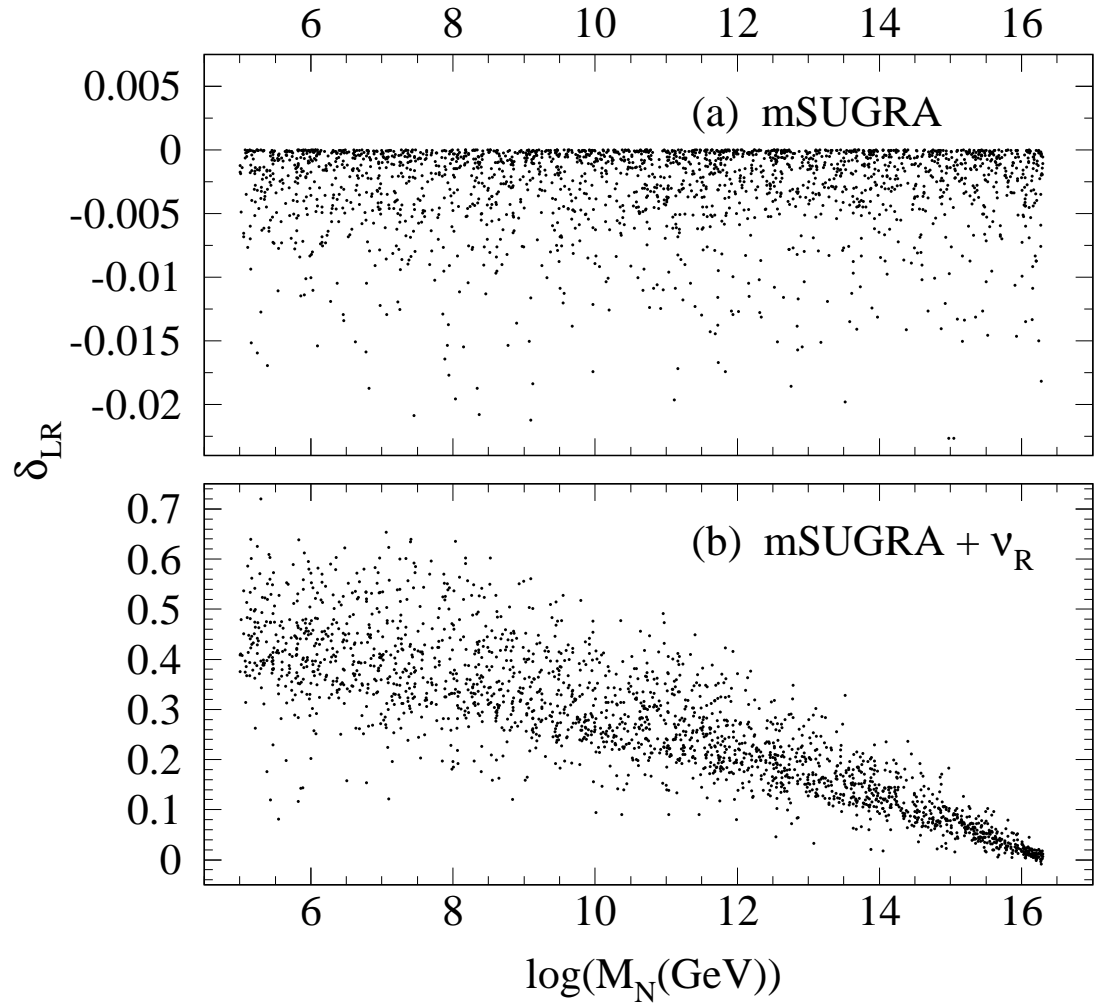}
\caption[]{The distribution of $\delta_{LR}$, defined in the text, for a
set of randomly generated ({\it a})~mSUGRA models, and ({\it b})~models
with a right handed neutrino, versus the $\nu_R$ mass $M_N$. In case
({\it a}) $M_N$ is randomly assigned. We assume that the top and
neutrino Yukawa couplings unify at the GUT scale.  }
\label{fig:delLR}
\end{figure}
\begin{figure}
\dofig{6in}{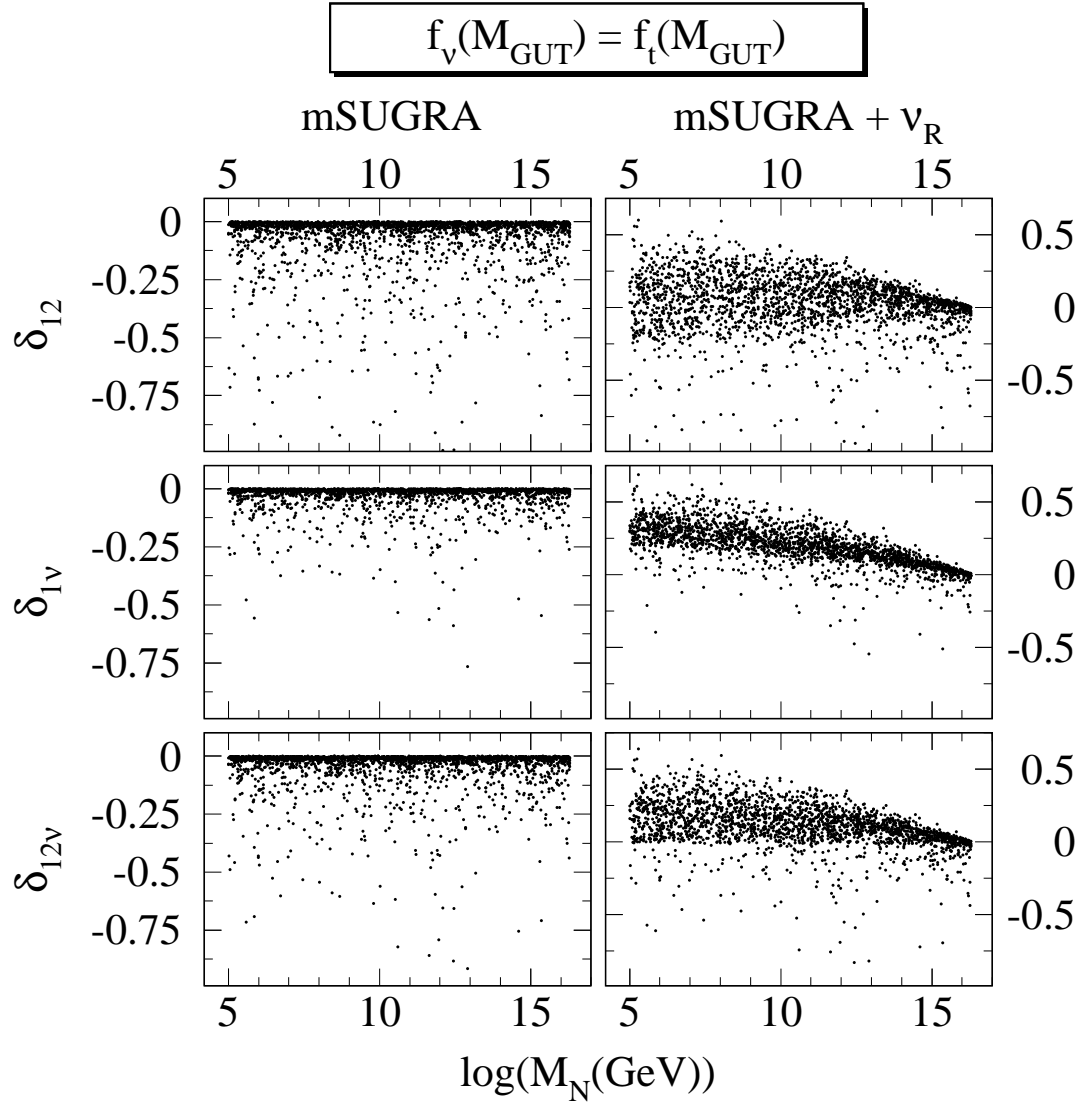}
\caption[]{The distribution of the quantities $\delta_{12}$ (first row),
$\delta_{1\nu}$ (second row) and $\delta_{12\nu}$, defined in Sec.~II of
the text, for the same set of models as in Fig.~\ref{fig:delLR} versus
$M_N$. The first column shows the results for mSUGRA models, and the
second one shows the corresponding results for models with a $\nu_R$. We
assume that the top and neutrino Yukawa couplings unify at the GUT
scale.}
\label{fig:mix}
\end{figure}
\begin{figure}
\dofig{6in}{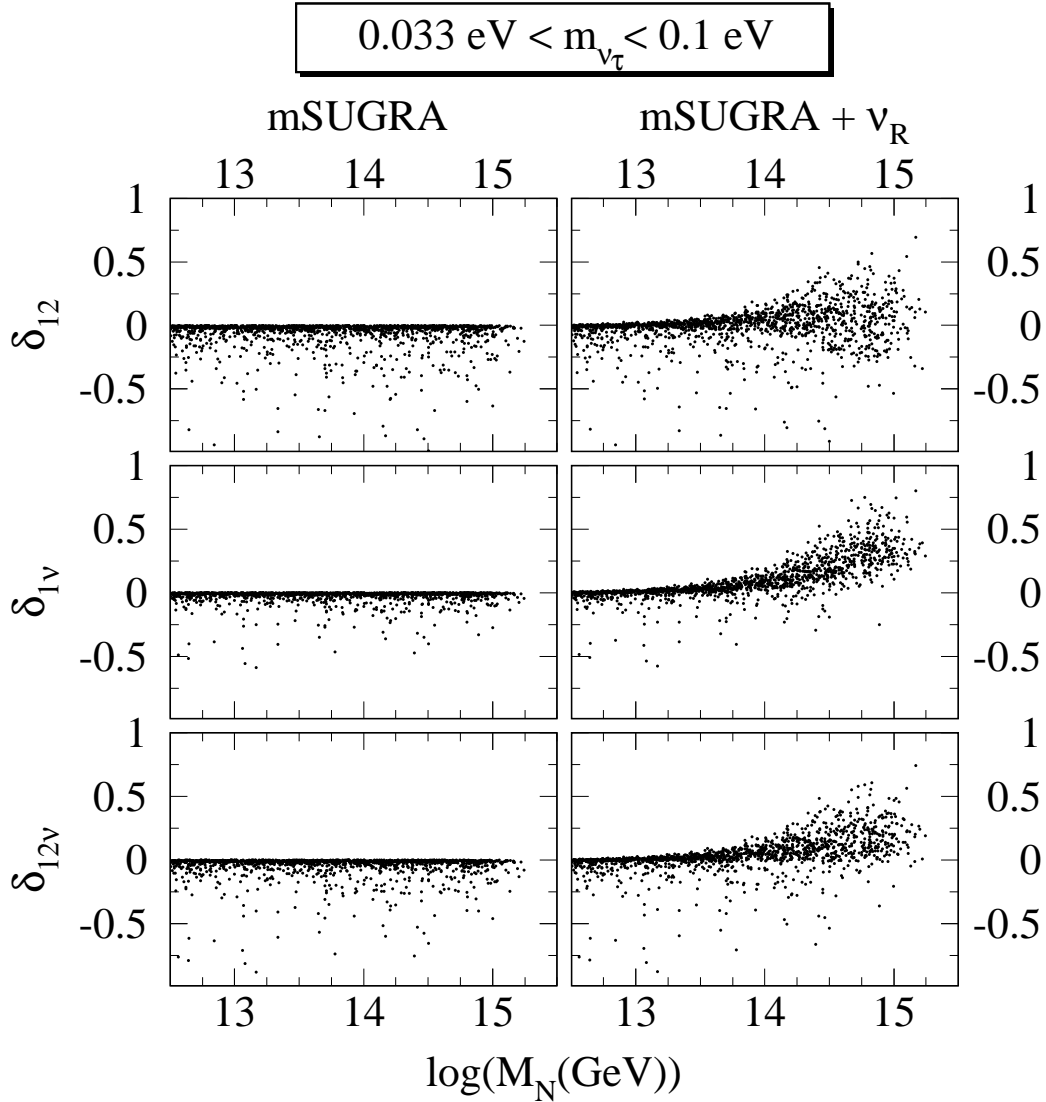}
\caption[]{The same as Fig.~\ref{fig:mix} except that instead of
assuming that the Yukawa couplings unify at the GUT scale, we require
that the mass of the tau neutrino is between the Super-Kamiokande range of
0.033~eV and 0.1~eV. }
\label{fig:neut}
\end{figure}
\begin{figure}
\dofig{6in}{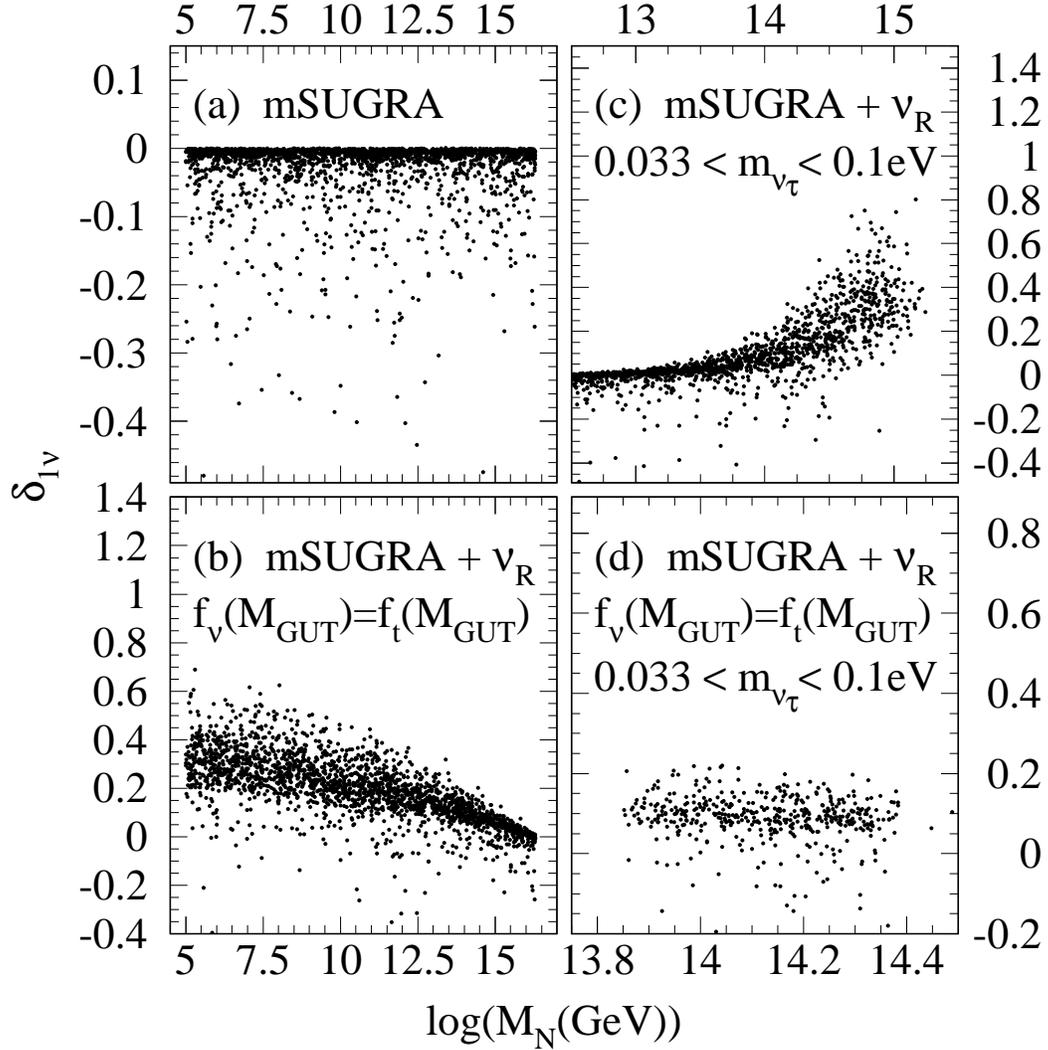}
\caption[]{The distribution of $\delta_{1\nu}$ versus the right handed
neutrino mass $M_N$ for the ({\it a})~mSUGRA model, ({\it b})~the RHN
model with unification of top and neutrino Yukawa couplings, ({\it
c})~the RHN model with the $\nu_{\tau}$ mass in the Super-Kamiokande
range, and ({\it d})~the RHN model with $f_t=f_{\nu_{\tau}}$ at the GUT
scale and the Super-Kamiokande constraint on $m_{\nu_{\tau}}$. Notice
the difference in the horizontal scales for the two columns.  }
\label{fig:nugut}
\end{figure}
\begin{figure}
\dofig{6in}{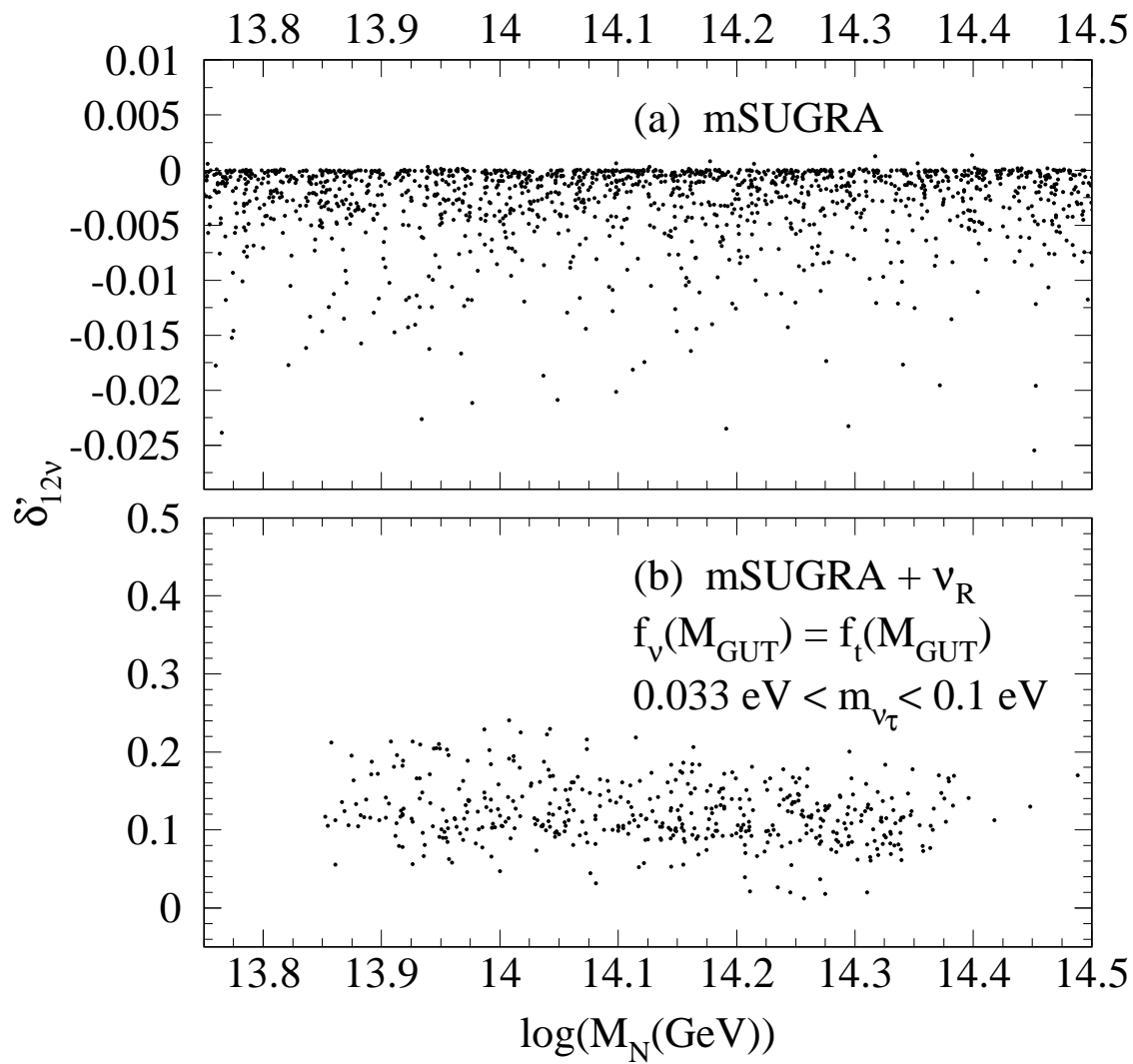}
\caption[]{The distribution of $\delta'_{12\nu}$ defined in the text
versus $M_N$ for the ({\it a})~mSUGRA model, and ({\it b})~the RHN model
with $f_t=f_{\nu_{\tau}}$ at the GUT
scale and the Super-Kamiokande constraint on $m_{\nu_{\tau}}$.}
\label{fig:clean}
\end{figure}
\begin{figure}
\dofig{6in}{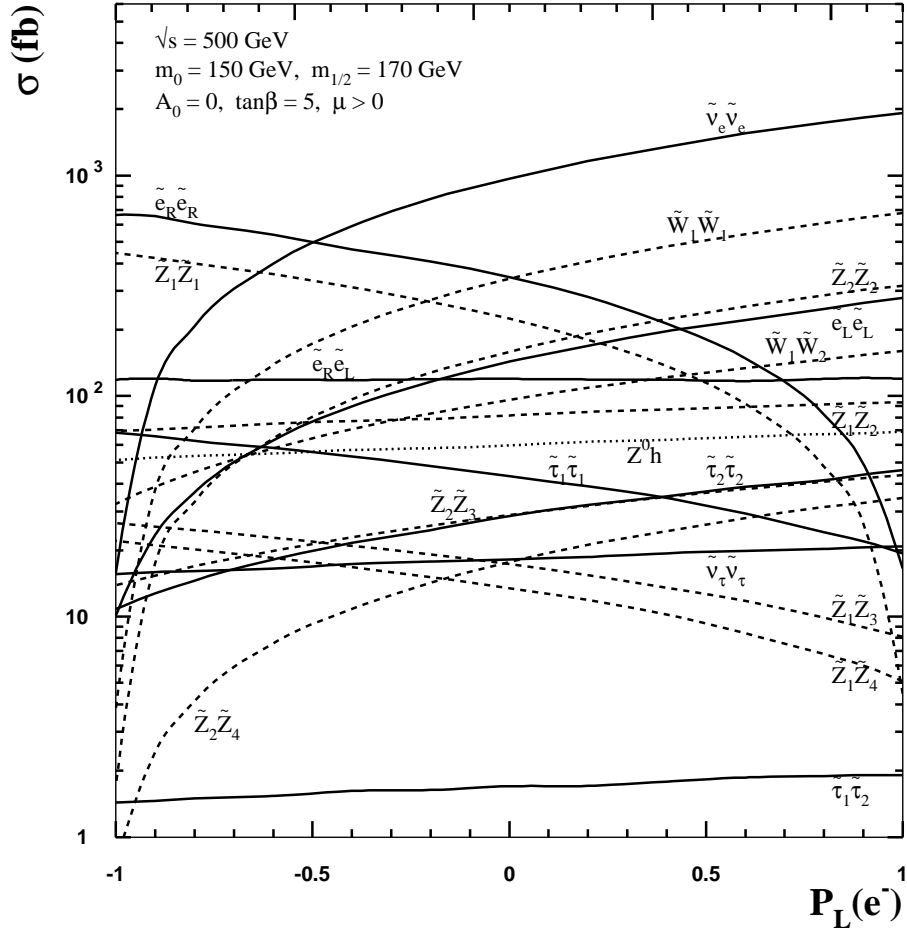}
\caption[]{Cross sections for various SUSY production
processes at an $e^+e^-$ collider with $\sqrt{s}=500$~GeV versus
the electron 
beam polarization parameter $P_L(e^-)$ for the case study in
Sec.~III. The solid lines show cross sections for sleptons, the dashed
lines for charginos and neutralinos and the dotted lines for Higgs boson
production mechanisms. The cross sections for $Ah$ and $ZH$ production
are below the 1~$fb$ level.}
\label{fig:csections}
\end{figure}
\begin{figure}
\dofig{6in}{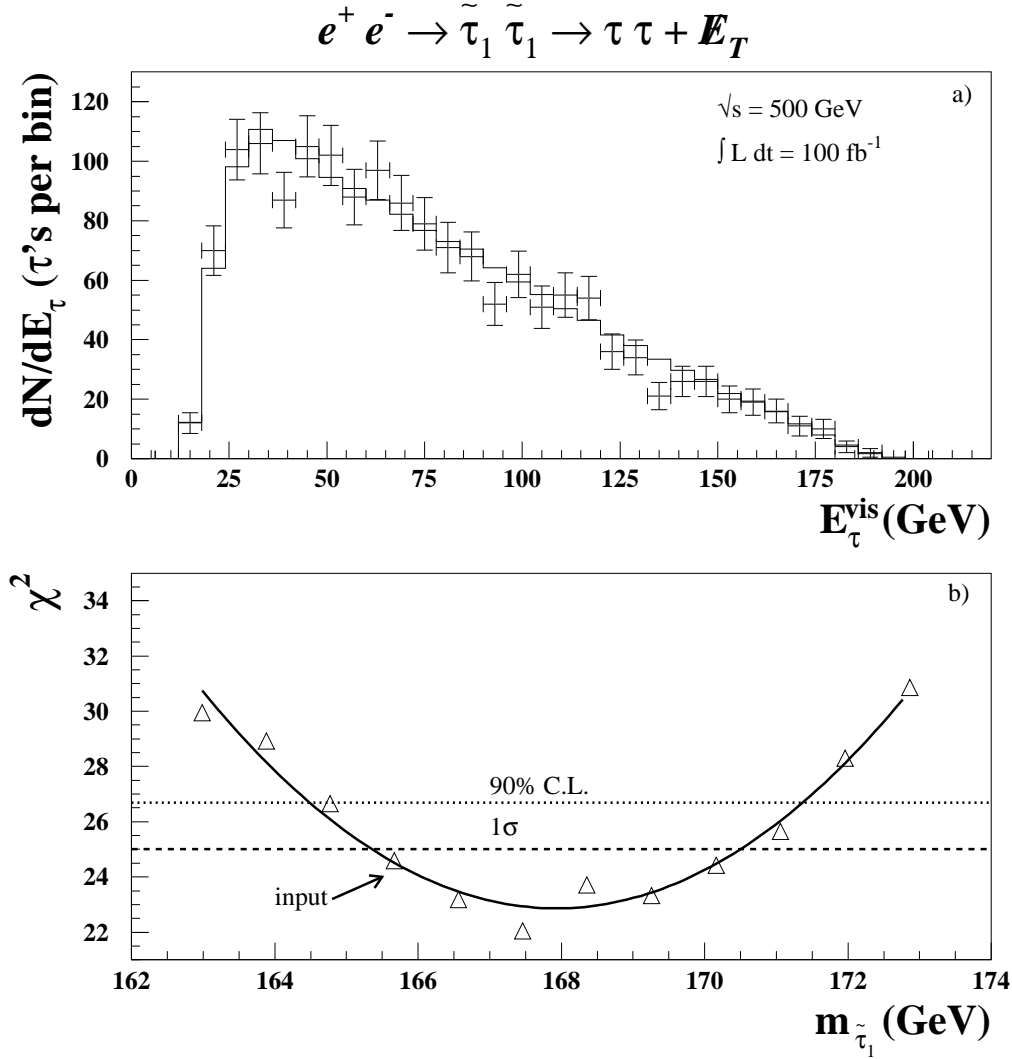}
\caption[]{({\it a})~The distribution of the visible energy from the
hadronic decays of taus produced via $e^+e^- \to \ttau_1\ttau_1 \to
\tau\tau +\eslt$ events at a 500~GeV collider for the case study of
Sec.~III. The solid histogram denotes the theoretical expectation after
cuts, while the points are the synthetic data for an integrated
luminosity of 100~$fb^{-1}$. Note that each event contributes two
visible taus. In ({\it b}), the values of $\chi^2$ obtained by a comparison
of synthetic data for several values of $m_{\ttau_1}$ with the theory
histogram in ({\it a}) above are shown by the triangles. The line is the
best-fit parabola through the triangles. The case study point is shown
denoted as input, and the dashed and dotted lines denote the $1\sigma$
and 90\% confidence levels for stau mass measurement. The integrated
luminosity is taken to be 100~$fb^{-1}$. }
\label{fig:tau1}
\end{figure}
\begin{figure}
\dofig{6in}{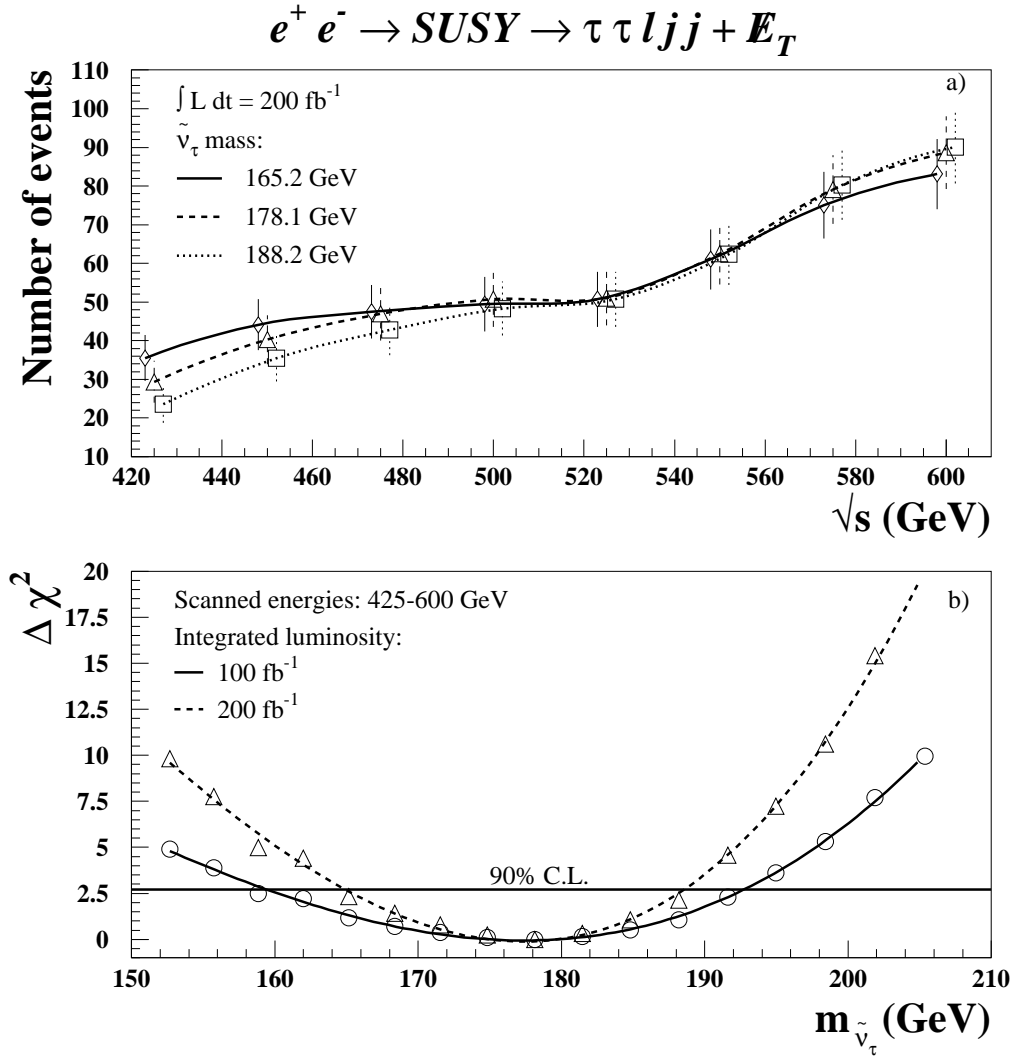}
\caption[]{({\it a}) The cross section after cuts described in the text
for $\tau\tau\ell jj+\eslt$ events from all SUSY processes (however, see
the text for a discussion of how SUSY backgrounds are computed) for the case
study point (dashed) and two other points with a lighter (solid) and
heavier (dotted) tau sneutrino versus the $e^+e^-$ center of mass
energy. The error bars reflect the statistical errors for an integrated
luminosity of 200~$fb^{-1}$. ({\it b})~The values of $\Delta\chi^2$ versus
$m_{\tnu_{\ttau}}$  for the
energy scan from 425~GeV to 600~GeV, assuming an integrated luminosity
of 100~$fb^{-1}$ (circles) and 200~$fb^{-1}$ (triangles). The curves are
a fit through these points. Also shown are the 90\% CL with which
$m_{\tnu_{\tau}}$ might be measurable. Here we have taken the test case
to be the ``data'' and only shown the change in $\chi^2$. }
\label{fig:snuall}
\end{figure}
\begin{figure}
\dofig{6in}{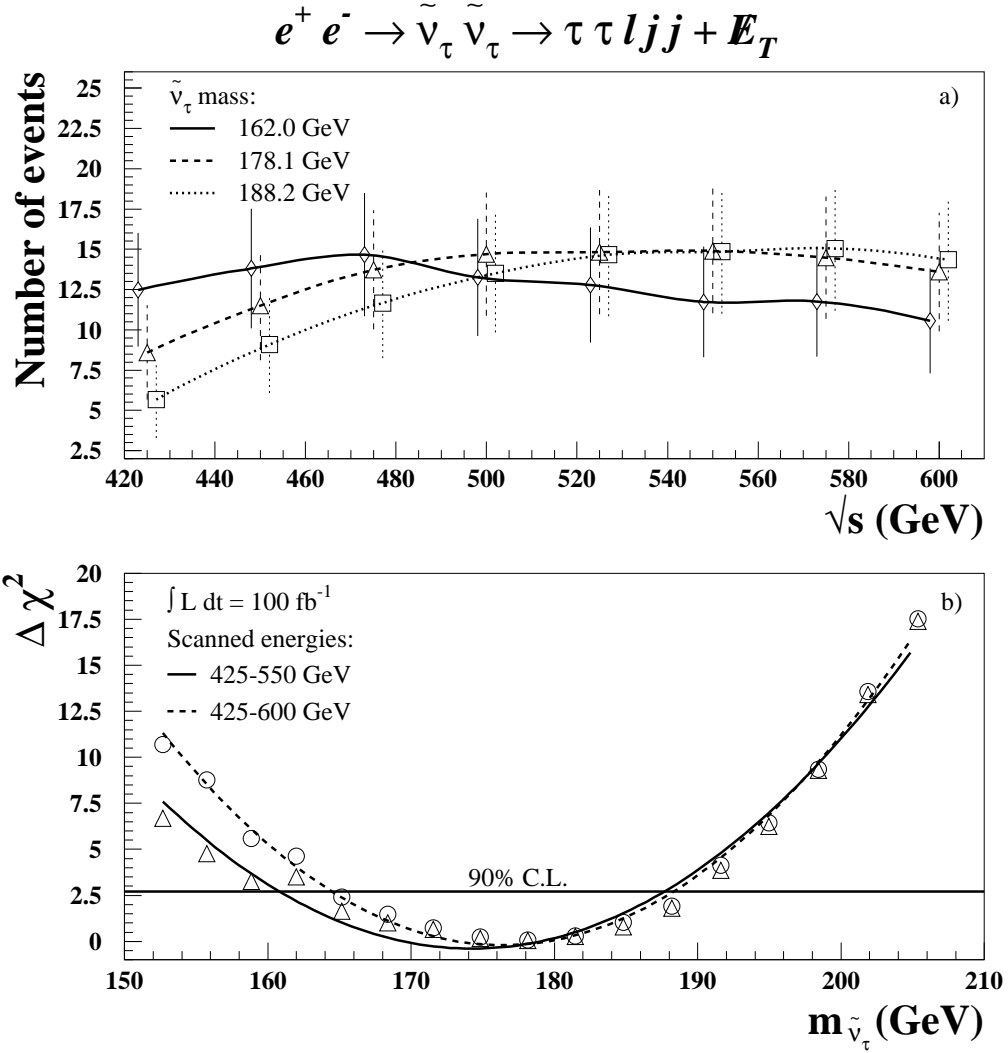}
\caption[]{The same as Fig.~\ref{fig:snuall} except that the SUSY
backgrounds are ignored and the integrated luminosity is fixed to be
100~$fb^{-1}$. Also in frame ({\it b}) we show the results for two sets
of energy scans: from 425~GeV to 550~GeV in steps of 25~GeV (triangles)
and from 425~GeV to 600~GeV (circles).  }
\label{fig:snuonly}
\end{figure}

\end{document}